\documentclass{article}
\usepackage{ijcai22}

\usepackage{times}
\usepackage{soul}
\usepackage{url}
\usepackage[hidelinks]{hyperref}
\usepackage[utf8]{inputenc}
\usepackage[small]{caption}
\usepackage{graphicx}
\usepackage{amsmath}
\usepackage{amsthm}
\usepackage{booktabs}
\usepackage{algorithm}
\usepackage{algorithmic}
\usepackage{bm}
\usepackage{enumitem}

\title{Poisoning Deep Learning Based Recommender Model\\in Federated Learning Scenarios}

\author{
Dazhong Rong
\and
Qinming He
\and
Jianhai Chen\thanks{Corresponding author}
\affiliations
College of Computer Science and Technology, Zhejiang University\\
\emails
\{rdz98, hqm, chenjh919\}@zju.edu.cn
}

\begin{document}

\maketitle

\begin{abstract}
Various attack methods against recommender systems have been proposed in the past years, and the security issues of recommender systems have drawn considerable attention.
Traditional attacks attempt to make target items recommended to as many users as possible by poisoning the training data.
Benifiting from the feature of protecting users' private data, federated recommendation can effectively defend such attacks.
Therefore, quite a few works have devoted themselves to developing federated recommender systems.
For proving current federated recommendation is still vulnerable, in this work we probe to design attack approaches targeting deep learning based recommender models in federated learning scenarios.
Specifically, our attacks generate poisoned gradients for manipulated malicious users to upload based on two strategies (\textit{i.e.,} random approximation and hard user mining).
Extensive experiments show that our well-designed attacks can effectively poison the target models, and the attack effectiveness sets the state-of-the-art.
\end{abstract}

\section{Introduction}
Recommender Systems (RS) have become one of the major channels for people to acquire information, and can directly influence people's perceptions while recommending items.
However, there are many vulnerabilities in RS which can be exploited to manipulate recommendation of items.
Attacks against RS aim to get target items recommended to as many users as possible by raising the predicted scores of target items.
One line of the attacks inject fake interactions to poison training data, termed as data poisoning attack~\cite{kapoor2017review}.
As attacker can only inject a limited number of fake interactions, these attacks are less effective.
Some works attempt to generate more crafted fake interactions by considering real interactions to improve attack effectiveness~\cite{li2016data,DBLP:conf/ndss/HuangMGL0X21}.
However, in Federated Learning (FL) scenarios, due to the fact that attacker has no access to real interactions, these works are not applicable.
Another line of the attacks directly poison recommender model by generating poisoned gradients, termed as model poisoning attack.
These attacks are specially targeted at FL scenarios and usually more effective than data poisoning attacks.
However, these attacks rely on attacker's prioir knowledge (\textit{e.g.,} items' popularity~\cite{zhang2021pipattack} or public interactions~\cite{rong2022fedrecattack}).
Under the circumstances that user privacy is strictly protected in FL scenarios, they also lose validity.
We summerized the key challenges here as following:
\begin{itemize}
	\item Existing attacks with attacker's prior knowledge are not applicable in FL scenarios.
	\item Existing attacks without attacker's prior knowledge have poor effectiveness.
\end{itemize}
We aim to poison deep learning based recommender model in FL scenarios without the prior knowledge.
To address the challenges above, we first approximate benign users' embedding vectors, and then generate poisoned gradients based on the approximated vectors rather than side information.

In this paper, we present two attack methods using different ways to approximate benign users' embedding vectors.
In the first attack method, inspired by recent work on self-supervised learning for recommendation~\cite{wu2021self,yao2021self} (See Section~\ref{A-ra} for more details), we approximate benign users' embedding vectors with gaussian distribution.
In the second attack method, inspired by OHEM~\cite{shrivastava2016training} (See Section~\ref{A-hum} for more details), we first initialize the approximated benign users' embedding vectors with gaussian distribution, and then optimize the vectors by gradient descent to mine hard users.
The key contributions of our work\footnote{https://github.com/rdz98/PoisonFedDLRS} are:
\begin{itemize}
	\item To the best of our knowledge, we are the first to present attacks against RS in FL scenarios without attacker's prior knowledge.
	\item We proposed two strategies (\textit{i.e.,} random approximation and hard user mining) to approximate benign users' embedding vectors in our attacks.
	Hard user mining can improve the attack effectiveness when the proportion of malicious users is extremely small.
	\item Extensive experiments on two real world datasets demonstrate that our attacks are effective and outperform all baseline attacks.
\end{itemize}

\section{Preliminaries}
In this section, we briefly introduce our base recommender model and the framework of federated recommendation.

\subsection{Base Recommender Model}
Neural Collaborative Filtering (NCF)~\cite{he2017neural} is one of the most widely used deep learning based recommender models, and has state-of-the-art performance of recommendation.
Without loss of generality, we adopt NCF as our base recommender model.

Respectively, Let $N$ and $M$ denote the number of users and items in the target recommender system.
Let $\bm{\mathcal{U}}$ and $\bm{\mathcal{I}}$ denote the set of $N$ users and the set of $M$ items.
Each user $\bm{u} \in \bm{\mathcal{U}}$ has an embedding vector $\bm{p}_u$ which describes its latent feature.
Similarly, each item $\bm{i} \in \bm{\mathcal{I}}$ has an embedding vector $\bm{q}_i$.
We use $\hat{Y}_{ui}$ to denote the predicted score between user $\bm{u}$ and item $\bm{i}$, which indicates the preference of user $\bm{u}$ for item $\bm{i}$.
In NCF, $\hat{Y}_{ui}$ is predicted as following:
\begin{equation}
	\hat{Y}_{ui} = \bm{\Upsilon} (\bm{p}_u, \bm{q}_i),
\end{equation}
where $\bm{\Upsilon}$ is the user-item interaction function. 
To model feedback signals with a high-level of non-linearities, NCF utilizes a Multi-Layer Perceptron (MLP) to learn the function $\bm{\Upsilon}$, as following:
\begin{equation}
	\bm{\Upsilon} (\bm{p}_u, \bm{q}_i) = \bm{\mathrm{a_{out}}} (\bm{h}^T \bm{\Phi} (\bm{p}_u \bm{\oplus} \bm{q}_i)),
\end{equation}
where $\bm{\mathrm{a_{out}}}$, $\bm{h}$, $\bm{\Phi}$ and $\bm{\oplus}$ denote the activation function, the weight vector, the MLP function and the vector concatenation respectively.
We use the sigmoid function as $\bm{\mathrm{a_{out}}}$ to restrict the model output to be in $(0,1)$.
More specifically, let $\bm{\mathrm{a_k}}$, $\bm{W}_k$ and $\bm{b}_k$ denote the activation function, the weight matrix and the bias vector in the $k$-th layer of the perceptron, respectively.
In NCF with $L$ hidden layers, the MLP function $\bm{\Phi}$ is as following:
\begin{equation}
	\bm{\Phi} (\bm{x}) = \bm{\mathrm{\phi_L}} (\dots (\bm{\mathrm{\phi_2}} (\bm{\mathrm{\phi_1}} (\bm{x})) \dots),
\end{equation}
where $\bm{\mathrm{\phi_k}} (\bm{x}) = \bm{\mathrm{a_k}} (\bm{W}_k^T \bm{x} + \bm{b}_k)$.
To achieve better performance of recommendation, we use Rectifier (ReLU) as the activation function $\bm{\mathrm{a_k}}$ in all perceptron layers. 

It is worth mentioning that, in federated learning scenarios, our attacks also apply to many other recommender models (\textit{e.g.,} NNCF~\cite{bai2017neural} and ONCF~\cite{DBLP:conf/ijcai/0001DWTTC18}) as long as they learn the user-item interaction function $\bm{\Upsilon}$ through deep neural networks.

\subsection{Framework of Federated Recommendation}
Our base recommender model is distributedly trained under the framework of federated recommendation.
More specifically, in FL scenarios, there is a central server and a large amount of individual user clients.
As each user corresponds to one of the user clients, we use user to represent its client for convenience.
Let $\bm{P}, \bm{Q}$ respectively denote the embedding matrices of users and items, where $\bm{p}_u$ is row of $\bm{P}$, and $\bm{q}_i$ is row of $\bm{Q}$.
Let $\bm{\mathcal{D}}_u$ denote the training dataset of user $\bm{u}$, which consists of item-score pairs $(\bm{i}, Y_{ui})$.
If user $\bm{u}$ has interacted with item $\bm{i}$, $Y_{ui}=1$ (\textit{i.e.,} a positive instance).
Otherwise, $Y_{ui}=0$ (\textit{i.e.,} a negative instance).
Note that, for each user $\bm{u}$, since there are a large amount of un-interacted items, the negative instances in $\bm{\mathcal{D}}_u$ are sampled with a ratio of $r:1$ as the number of negative instances against that of positive.
Let $\bm{\Theta}$ denote all trainable model parameters except for users' embedding matrix $\bm{P}$
(In NCF, $\bm{\Theta}=\{\bm{Q}, \bm{h}, \bm{W}_1, \bm{b}_1, \bm{W}_2, \bm{b}_2, \dots, \bm{W}_L, \bm{b}_L\}$).
For each user $\bm{u}$, $\bm{p}_u$ and $\bm{\mathcal{D}}_u$ are stored locally on its own device.
Meanwhile, $\bm{\Theta}$ are stored on the central server.

To train our base recommender model, we adopt the Binary Cross-Entropy (BCE) loss to quantify the difference between the model predicted scores and the ground-truth scores on training dataset.
Let $\bm{\mathcal{L}}_u$ denote the loss function for user $u$.
We can consider $\bm{\mathcal{L}}_u$ as a function of $\bm{p}_u$ and $\bm{\Theta}$ as following:
\begin{equation}
	\bm{\mathcal{L}}_u = -\sum_{(\bm{i}, Y_{ui}) \in \bm{\mathcal{D}}_u} Y_{ui}\log \hat{Y}_{ui}+(1-Y_{ui})\log (1-\hat{Y}_{ui}).
\end{equation}
Once again note that, without loss of generality, our attacks are valid while other popular loss functions (\textit{e.g.,} BPR loss~\cite{DBLP:conf/uai/RendleFGS09}) are adopted.

In every training round, the central server randomly selects a subset of users to participate in training.
Let $\bm{\mathcal{U}}^t$ denote the selected user subset in the $t$-th round.
Let $\bm{p}_u^t$ and $\bm{\Theta}^t$ denote $\bm{p}_u$ and $\bm{\Theta}$ in the $t$-th round, respectively.
In the $t$-th round, the central server first sends a copy of $\bm{\Theta}^t$ to each selected users in $\bm{\Theta}^t$.
Then, for each selected user $\bm{u}$, with $\bm{\mathcal{L}}_u$ computed, it derives the gradients of $\bm{p}_u^t$ and $\bm{\Theta}^t$ denoted by $\nabla \bm{p}_u^t$ and $\nabla \bm{\Theta}_u^t$.
Moreover, each selected user $\bm{u}$ updates $\bm{p}_u$ locally with learning rate $\eta$ as following:
\begin{equation}
	\bm{p}_u^{t+1}=\bm{p}_u^t - \eta\nabla \bm{p}_u^t,
\end{equation}
and uploads $\nabla \bm{\Theta}_u^t$ to the central server.
At last, after the central server collects all uploaded gradients from selected users, it updates $\bm{\Theta}$ by aggregating the uploaded gradients as following:
\begin{equation}
	\bm{\Theta}^{t+1}=\bm{\Theta}^t-\eta\sum_{\bm{u}\in \bm{\mathcal{U}}^t}\nabla \bm{\Theta}_u^t.
\end{equation}
There are repetitive rounds of training until the recommender model converges.

\section{Our Attacks}\label{our-attacks}
In order to avoid ambiguity, we use $\bm{\mathcal{U}}$ and $\bm{\tilde{\mathcal{U}}}$ to denote the set of benign users and the set of malicious users injected by attacker, respectively.
We assume that for each user the recommender model recommends $K$ items with the highest predicted scores among the items that the user did not interact with.
Let $\bm{\mathcal{I}}_u$ denote the top-K recommended items for user $\bm{u}$.
The Exposure Ratio at rank $K$ (ER@K) of item $\bm{i}$ is defined as: $|\{u\in \bm{\mathcal{U}} | i \in \bm{\mathcal{I}}_u\}| / |\{u\in \bm{\mathcal{U}} | (\bm{i}, 1) \notin \bm{\mathcal{D}}_u\}|$.

Let $\bm{\tilde{\mathcal{I}}}$ denote the set of target items.
Let $\bm{\varepsilon_i}$ denote ER@K of item $\bm{i}$.
The goal of our attacks is to raise $\text{ER@K}$ of target items (\textit{i.e.,} maximize $\sum_{i\in \bm{\tilde{\mathcal{I}}}} {\bm{\varepsilon_i}}$).
To achieve the goal, attacker manipulates malicious users to upload well-crafted poisoned gradients to central server.
In this section, we propose two attack methods A-ra and A-hum that use different strategies (\textit{i.e.,} random approximation and hard user mining) to generate the poisoned gradients.
In the following, we will introduce them respectively.

\subsection{Attack with Random Approximation (A-ra)}\label{A-ra}
The attacks against RS in FL scenarios can be considered as a special class of backdoor attacks.
Gu \textit{et al.}~\shortcite{gu2017badnets} backdoored neural networks in the outsourced training scenario by mixing clean inputs and backdoored inputs.
Inspired by the previous work, we backdoor the target recommender model similarly by uploading a mix of clean gradients and poisoned gradients.
Let $\bm{\tilde{\mathcal{L}}}$ denote the loss function which can indicate the goal of our attacks.
In the $t$-th round of training, each selected benign user $\bm{u}$ uploads gradients of $\bm{\mathcal{L}}_u$ (\textit{i.e.,} clean gradients).
To achieve the goal of our attacks, we manipulate the selected malicious users to upload gradients of $\bm{\tilde{\mathcal{L}}}$ (\textit{i.e.,} poisoned gradients).
The actual loss function of the recommender model under our attacks can be represented as $\sum_{\bm{u} \in \bm{\mathcal{U}}}\bm{\mathcal{L}}_u + \alpha \bm{\tilde{\mathcal{L}}}$,
where $\alpha$ is a positive coefficient that trades off between the model validity and the attack effectiveness.
Note that attacker can adjust $\alpha$ by changing the learning rate for malicious users.

Unfortunately, we can not bring $\text{ER@K}$ of target items into $\bm{\tilde{\mathcal{L}}}$, because:
\begin{itemize}
	\item ER@K is a highly non-differentiable discontinuous function of $\bm{\Theta}$, hence leads to the difficulty in the computation of effective poisoned gradients.
	
	\item In our attacks, each benign user's embedding vector and training dataset are unknown to attacker.
	However, ER@K depends on these parameters.
\end{itemize}
To address the problems, we adopt two steps to design a proper loss function $\bm{\tilde{\mathcal{L}}}$ for our attacks.

\textbf{Step 1: Using predicted scores to indicate $\text{ER@K}$.}\quad
For each item $\bm{i}$, there is a positive correlation between its $\text{ER@K}$ (\textit{i.e.,} $\bm{\varepsilon_i}$) and its average predicted score to all benign users (\textit{i.e.,} $\frac{1}{|\bm{\mathcal{U}}|} \sum_{\bm{u}\in \bm{\mathcal{U}}} \hat{Y}_{ui}$).
Instead of maximizing $\sum_{i\in \bm{\tilde{\mathcal{I}}}} {\bm{\varepsilon_i}}$, we can change the goal of our attacks to maximizing $\frac{1}{|\bm{\mathcal{U}}|}\sum_{i\in \bm{\tilde{\mathcal{I}}}} \sum_{u\in \bm{\mathcal{U}}} {\hat{Y}_{ui}}$.
Different from $\bm{\varepsilon_i}$, $\hat{Y}_{ui}$ is a continuous and differentiable function of $\bm{\Theta}$.
However, if user $u$ have interacted with the target item $\bm{i}$, $\hat{Y}_{ui}$ does not affect $\bm{\varepsilon_i}$.
Moreover, if the target item $\bm{i}$ has already been in user $\bm{u}$'s top-$K$ recommendation list, increasing $\hat{Y}_{ui}$ is meaningless.
Due to the above two reasons, our new goal leads to a little loss of precision.

\textbf{Step 2: Approximating users' embedding vectors with normal distribution.}\quad
The scores predicted by target recommender model depend on $\bm{P}$ and $\bm{\Theta}$.
In the $t$-th round of training, the central server sends a copy of $\bm{\Theta}^t$ to each user in $\bm{\mathcal{U}}^t$.
Attacker can easily access $\bm{\Theta}^t$ through the malicious users in $\bm{\mathcal{U}}^t$.
However, $\bm{P}$ is always inaccessible for attacker, because $\bm{P}$ is distributedly stored on benign users' clients as their embedding vectors.
To carry out our attacks, we have to approximate $\bm{P}$.
Some recent studies on self-supervised learning for recommendation~\cite{wu2021self,yao2021self} introduce contrastive learning to make the unit embedding vectors of different users or different items far apart, so as to improve the accuracy and robustness for recommendation.
Inspired by these studies, we realize that the directions of users' embedding vectors are expected to be uniformly distributed.
With the additional consideration of the effect of L2-regularization, we roughly assume that users' embedding vectors follow a gaussian distribution.
Therefore, we approximate benign user's embedding vector as a vector-valued random variable $\bm{\hat{\mathrm{p}}} \sim N(0,\sigma^2)$, where $\sigma$ is a hyperparameter.
Let $\bm{\hat{\mathrm{p}}}_j$ denotes our $j$-th approximated embedding vector.
We further change the goal of our attack to maximizing $\frac{1}{n} \sum_{i\in \bm{\tilde{\mathcal{I}}}} {\sum_{j=1}^n {\bm{\Upsilon} (\bm{\hat{\mathrm{p}}}_j, \bm{q}_i)}}$, where $n$ is a hyperparameter which indicates the stability of our attack.
The larger $n$ is, the slighter the effect of randomness to our attack is.

Considering the stealthiness of our attacks, $\bm{\tilde{\mathcal{L}}}$ is designed as following:
\begin{equation}
	\bm{\tilde{\mathcal{L}}} (\bm{\Theta}) = \frac{1}{n} \sum_{i\in \bm{\tilde{\mathcal{I}}}} {\sum_{j=1}^n {-\log \bm{\Upsilon} (\bm{\hat{\mathrm{p}}}_j, \bm{q}_i)}}.
\end{equation}
In order to make $\bm{\tilde{\mathcal{L}}}$ have a similar structure to $\bm{\mathcal{L}}$, we maximize $\bm{\Upsilon} (\bm{\hat{\mathrm{p}}}_j, \bm{q}_i)$ by minimizing the BCE loss between $\bm{\Upsilon} (\bm{\hat{\mathrm{p}}}_j, \bm{q}_i)$ and the maximum score (\textit{i.e.,} $1$).

To summarize the flow of our attacks in the $t$-th round of training, the selected malicious users first derive the gradients $\widetilde{\nabla \bm{\Theta}^t}$ of $\bm{\Theta}^t$ with $\bm{\tilde{\mathcal{L}}}$ computed, and then upload $\widetilde{\nabla \bm{\Theta}^t}$ to the central server.

\subsection{Attack with Hard User Mining (A-hum)}\label{A-hum}
In A-ra, the predicted scores of target items to most benign users will increase significantly under the effect of poisoned gradients.
However, for the benign user $\bm{u}$ who takes the target item $\bm{i}$ as a negative instance (\textit{i.e.,} $(i,0)\in\bm{\mathcal{D}}_u$), its uploaded gradients have opposite effect to poisoned gradients on the predicted score $\hat{Y}_{ui}$.
For convenience, we call the set of such benign users as hard users of item $\bm{i}$.
It is difficult to productively increase the predicted scores of target items to their corresponding hard users.
As a result, our attack effectiveness is limited.

We argue that Step 2 in A-ra is still improvable. 
Shrivastava \textit{et al.}~\shortcite{shrivastava2016training} introduced an Online Hard Example Mining (OHEM) algorithm which automatically select hard examples to train region-based ConvNet detectors in a more effective and efficient way.
Inspired by OHEM, to address the above challenge and further improve our attack effectiveness, we propose A-hum.
In A-hum, we use gradient descent to mine hard users, and generate poisoned gradients in allusion to them.
More specifically, the flow of A-hum in the $t$-th round of training is as following:

\begin{enumerate}
	\item For each target item $\bm{i}$, we first initialize the approximated embedding vectors of its hard user as $\bm{\hat{\mathrm{p}}} \sim N(0,\sigma^2)$, which is the same as that in Step 2 in A-ra.
	Let $\bm{\hat{\mathrm{p}}}^i_j$ denote the $j$-th approximated embedding vector for the target item $\bm{i}$.
	
	\item For each target item $\bm{i}$ and its hard user $\bm{u}$, due to the fact that $-\log(1-\bm{\Upsilon} (\bm{p}_u, \bm{q}_i))$ is one term of $\bm{\mathcal{L}}_u$, its value is expected to be small.
	Therefore, we define the loss function of hard user mining for the target item $\bm{i}$ as following:
	\begin{equation}
		\bm{l}_i (\bm{p}) = -\log(1-\bm{\Upsilon} (\bm{p}, \bm{q}_i)).
	\end{equation}
	And then for each $\bm{\hat{\mathrm{p}}}^i_j$, we optimize it to minimize $\bm{l}_i (\bm{\hat{\mathrm{p}}}^i_j)$ by gradient descent as following:
	\begin{equation}
		\bm{\hat{\mathrm{p}}}^i_j \leftarrow \bm{\hat{\mathrm{p}}}^i_j - \xi\frac{\partial \bm{l}_i}{\partial \bm{\hat{\mathrm{p}}}^i_j} (\bm{\hat{\mathrm{p}}}^i_j),
	\end{equation}
	where $\xi$ is the learning rate for hard user mining.
	This optimization process is repeated for $\beta$ iterations, where $\beta$ is a hyperparameter.
	
	\item Finally, we design $\bm{\tilde{\mathcal{L}}}$ as following:
	\begin{equation}
		\bm{\tilde{\mathcal{L}}} (\bm{\Theta}) = \frac{1}{n} \sum_{i\in \bm{\tilde{\mathcal{I}}}} {\sum_{j=1}^n {-\log \bm{\Upsilon} (\bm{\hat{\mathrm{p}}}_j^i, \bm{q}_i)}},
	\end{equation}
	which specifically aims to solve the difficulty brought by the hard users.
	As in A-ra, the selected malicious users first derive the gradients $\widetilde{\nabla \bm{\Theta}^t}$ of $\bm{\Theta}^t$ with $\bm{\tilde{\mathcal{L}}}$ computed, and then upload $\widetilde{\nabla \bm{\Theta}^t}$ to the central server.
\end{enumerate}


\section{Related Work}
In Federated Recommendation (FR), each user's privacy (historical interactions) is kept locally on the user client itself instead of being collected and uploaded to the central server, which means that the user's privacy will not be exposed to anyone else including the central server.
Because users' privacy is well protected, FR becomes a hot topic of research, and a large number of studies (\textit{e.g.,}~\cite{ammad2019federated,lin2020fedrec,liang2021fedrec++}) spring up on this research area.
Fueled by the maturity of FR techniques, the recommender systems based on FR are becoming continuously more widely used in diverse scenarios, as long as the security issues of FR are emerging. 
Although some research has been carried out on FR, few studies have investigated the security of FR.
In the following we will discuss the studies related to the attacks against FR from three perspectives respectively.

\textbf{Attacks against centralized recommendation.}\quad
Data poisoning attack is the main stream of attacks against centralized recommendation, originating from adversarial machine learning.
In such attacks, attacker injects a small amount of fake users and generates well-crafted interactions by manipulating the fake users to interact with certain specific items.
The recommender model trained on the poisoned data (fake users' interactions) will predict abnormally high scores of target items to many users.
Kapoor \textit{et al.}~\shortcite{kapoor2017review} introduced two naive data poisoning attacks (\textit{i.e.,} random attack and bandwagon attack).
In random attack, fake users interact with target items and randomly selected items under attacker's manipulation.
In bandwagon attack, fake users interact with target items and certain items with high popularity under attacker's manipulation.
Note that bandwagon attack relies on side information about items' popularity.
Both random attacks and bandwagon attacks don't take into consideration the algorithm adopted by the target recommender model.
As a result, these two attacks have limited effectiveness.
To improve on this shortcoming, Li \textit{et al.}~\shortcite{li2016data}, Fang \textit{et al.}~\shortcite{fang2018poisoning} and Huang \textit{et al.}~\shortcite{DBLP:conf/ndss/HuangMGL0X21} developed data poisoning attacks for matrix factorization based, graph based and deep learning based recommender systems respectively.
Note that all these three attacks rely on attacker's prior knowledge of historical interactions.
Even though these attacks consider the recommendation algorithm and take use of attacker's prior knowledge, their effectiveness is still limited as attacker can only poison a minor part of the training data.
Unlike centralized recommendation, in FR, attacker can directly poison the gradients uploaded to the central server through malicious clients.
Hence, theoretically, attacks against FR can achieve better effectiveness than attacks against centralized recommendation.
More importantly, most of data poisoning attacks require access to all or major part of user-item interactions.
These attacks are quite impractical in FR scenarios, because each user's historical interactions are kept locally and attacker has no access to them.

\textbf{Attacks against federated learning.}\quad
McMahan \textit{et al.}~\shortcite{mcmahan2017communication} proposed a framework of Federated Learning (FL), which trains deep neural network distributedly with thousands participants.
In every training round of FL, the central server randomly selects a subset of user clients, then share the master model with these clients.
Each of these clients trains the master model locally with the training data kept on the client itself, then uploads the gradients of model parameters to the central server.
At the end of the round, the central server updates the master model by averaging the uploaded gradients.
The issues of security arise along with the rapid development of FL.
A large number of attacks against FL are proposed.
Bagdasaryan \textit{et al.}~\shortcite{bagdasaryan2020backdoor} introduced model poisoning attacks to which FL is generically vulnerable.
In model poisoning attack, attacker directly influences master model by controlling malicious clients to upload poisoned gradients.
Subsequently, Fang \textit{et al.}~\shortcite{fang2020local} performed the first systematic study on attacking Byzantine-robust FL and proposed local model poisoning attacks against Byzantine-robust FL.
However, these attacks against FL rely on the complete access to the model, which is not realistic in FR.
FR is a special case of FL, because in FR users' embedding vectors are kept locally and no one has the complete access to the model.
Therefore, these attacks against FL do not work in FR.

\textbf{Attacks aginst federated recommendation.}\quad
Previous studies on the attacks against centralized recommendation and FL are not effective in FR.
Moreover, there are few studies on the attacks especially designed against FR.
Zhang \textit{et al.}~\shortcite{zhang2021pipattack} and Rong \textit{et al.}~\shortcite{rong2022fedrecattack} presented model poisoning attacks to FR named PipAttack and FedRecAttack respectively.
However, both two attacks require attacker to have prior knowledge (\textit{i.e.,} the side information that reflects each item's popularity or some public user-item interactions), which is not generic in all FR scenarios.
From the above discussion we can conclude that, although there is increasing concern about the security issues of FR, the attacks against FR are still under-explored.


\section{Experiments}
\begin{figure*}[t]
	\centering
	\includegraphics[width=0.99\linewidth]{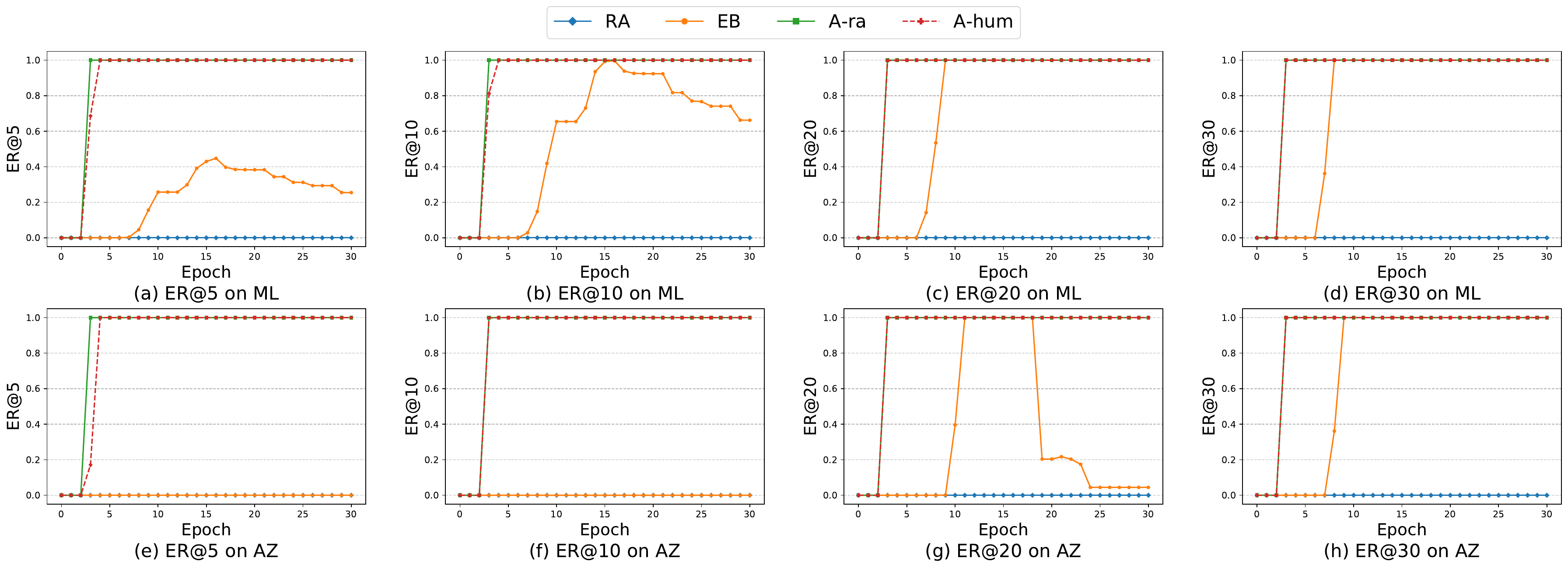}
	\caption{Attack Effectiveness under Different Attacks on ML ((a)-(d)) and AZ ((e)-(h)).}
	\label{f1}
\end{figure*}
\begin{figure*}[t]
	\centering
	\includegraphics[width=0.99\linewidth]{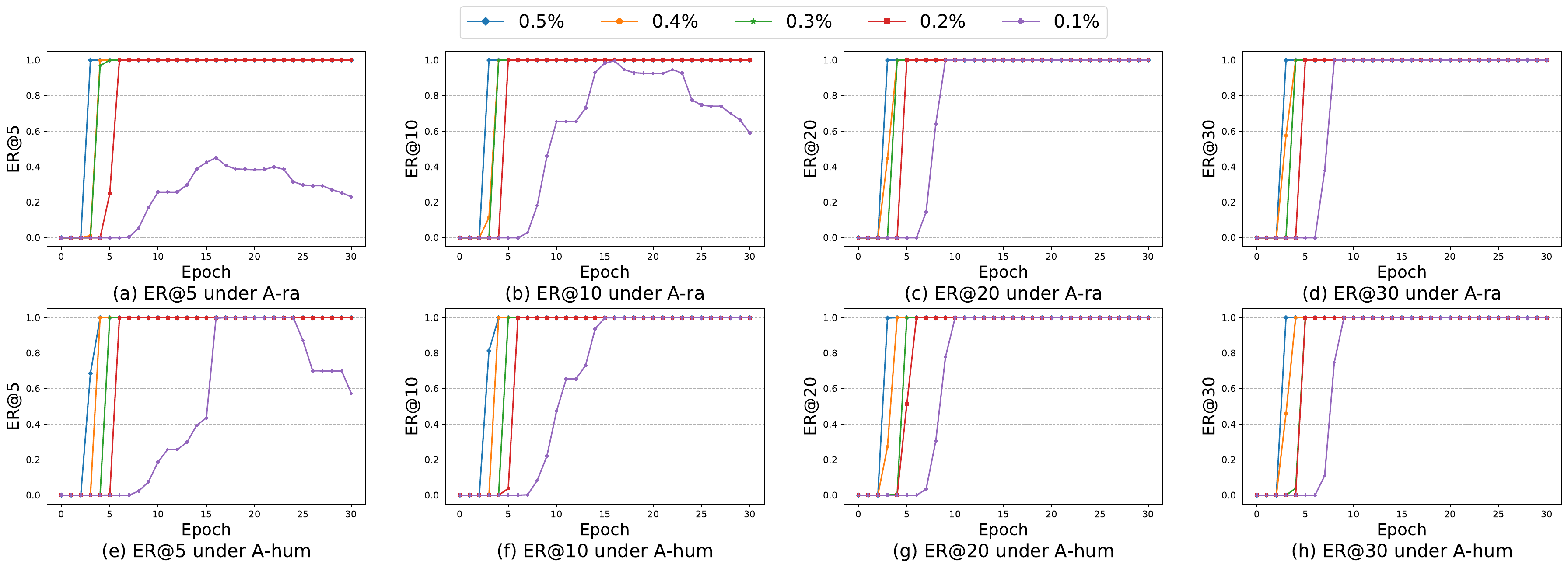}
	\caption{Attack Effectiveness under A-ra ((a)-(d)) and A-hum ((e)-(h)) with Different Malicious User Proportions.}
	\label{f2}
\end{figure*}
In this section, we conduct extensive experiments to address the following research questions (RQs):
\begin{itemize}
	\item \textbf{RQ1}: How is the effectiveness of our attacks compared to that of existing attacks?
	\item \textbf{RQ2}: Are our proposed attacks still effective when the proportion of malicious users is extremely small?
	\item \textbf{RQ3}: Does hard user mining really improve the attack effectiveness?
\end{itemize}

\subsection{Experimental Settings}
\textbf{Dataset.}\quad
We experiment with two popular and publicly accessible datasets: \textbf{MovieLens (ML)} and \textbf{Amazon Digital Music (AZ)}.
For ML, we use the version containing 1,000,208 ratings involving 6,040 users and 3,706 movies, where each user has at least 20 ratings.
For AZ, we use the version containing 169,781 reviews involving 16,566 users and 11,797 products of digital music, where each user and product has at least 5 reviews.
In both datasets, we convert the user-item interactions (\textit{i.e.,} ratings and reviews) into implicit data following \cite{he2017neural}, and divide each user's interactions into training set and test set in the ratio of $4:1$.
Note that we intentionally used two datasets of completely different sizes in order to validate the robustness of our attacks.

\textbf{Evaluation Protocols.}\quad
We assume that malicious users are injected by attacker before the training starts.
As malicious users upload poisoned gradients in each epoch of training, we evaluate effectiveness of our attacks after each epoch.
We adopt average Exposure Ratio at $K$ (\textbf{ER@K}) of target items as our metric, where $\text{ER@K}$ is defined in Section~\ref{our-attacks}. 
High metric values indicate strong attack effectiveness.
Note that to ensure fairness, we select $T$ most unpopular items (\textit{i.e.,} the items with the least number of interactions) as our target items on both datasets.
The average ER@K of these items is the most difficult to be boosted.
In our experiments, to observe the attack effects from various scales, we set $K=5,10,20,30$ respectively.

\textbf{Baseline Attacks.}\quad
Following the common setting for federated recommendation, we assume that attacker does not have any prior knowledge at all.
Bandwagon attack~\cite{kapoor2017review,gunes2014shilling}, PipAttack~\cite{zhang2021pipattack} and FedRecAttack~\cite{rong2022fedrecattack} rely on side infomation of items' popularity or public interactions, hence they are not applicable under such circumstances.
The data poisoning attacks in~\cite{li2016data,DBLP:conf/ndss/HuangMGL0X21} rely on historical interactions of benign users, they are either not applicable. 
We choose the attacks which are still practical under such circumstances as our baseline attacks:
\textbf{i) Random Attack (RA)}~\cite{kapoor2017review}.
It injects fake users as malicious users, and manipulates them to interact with both target items and randomly selected items.
\textbf{ii) Explicit Boosting (EB)}~\cite{zhang2021pipattack}.
It is one component of PipAttack which does not rely on attacker's prior knowledge.
It explicitly boosts the predicted scores of target items for malicious users.
More specifically, different from benign users, malicious users take the explicit item promotion objective as their loss function.
Other than that, malicious users behave same as benign users.

\textbf{Parameter Settings.}\quad
We adopt NCF with 2 hidden layers as our base recommender model. 
The dimensions of both hidden layer are set to 8.  
The dimensions of users' embedding vectors and items' embedding vectors are also set to 8.
The learning rate $\eta$ for both benign users and malicious users is set to $0.001$.
The base recommender model is federated trained for $30$ epochs on both ML and AZ to ensure convergence for recommendation.
Moreover, we set $r$, $T$, $n$, $\sigma$, $\xi$ and $\beta$ to $4$, $1$, $10$, $0.01$, $0.001$ and $30$, respectively.
Let $\rho$ denote the proportion of malicious users.
Unless otherwise mentioned, $\rho$ is set to $0.5\%$ on ML and $0.1\%$ on AZ.

\subsection{Comparison of Attack Effectiveness (RQ1)}
We compare the effectiveness of baseline attacks and ours on ML and AZ.
We set different proportion of malicious users on two datasets because attacks are more likely to achieve stronger effectiveness on AZ than on ML.
More specifically, we set $\rho=0.5\%$ for ML and set $\rho=0.1\%$ for AZ.
Figure~\ref{f1} shows ER@5, ER@10, ER@20 and ER@30 under baseline attacks and our attacks on ML and AZ.
As the figure shown:
\textbf{(i)} Due to the proportion of malicious users is too small for the naive data poisoning attack RA, it has no effects.
\textbf{(ii)} Though EB creates slight effects, its effectiveness is unstable.
It could be easily seen that the results of EB continuously descrease once the training process passed half.
\textbf{(iii)} Our proposed A-ra and A-hum outperforms baseline attacks consistently in terms of all metrics.
The metric values of both A-ra and A-hum increase rapidly to $1$ in the first few epochs, maintaining in high level throughout the whole rest of the training epochs.

\subsection{Impact of Malicious User Proportion (RQ2)}
To further explore the effectiveness of our attacks under certain circumstances, we evaluate our attacks with smaller proportion of malicious users on ML.
Figure~\ref{f2} shows attack effectiveness of A-ra and A-hum with $\rho=0.5\%, 0.4\%, 0.3\%, 0.2\%, 0.1\%$ on ML.
As the figure shown, the effectiveness of both attacks does not drop until the proportion of malicious users is decreased to $0.1\%$.
Note that our attacks demand the lowest proportion of malicious users compared to any existing attacks against recommender systems (including the attacks with attacker's prior knowledge), not to mention that most attacks are only effective when the proportion of malicious users is more than $5\%$.

\subsection{Ablation Test (RQ3)}
\begin{figure}[t]
	\centering
	\includegraphics[width=0.99\linewidth]{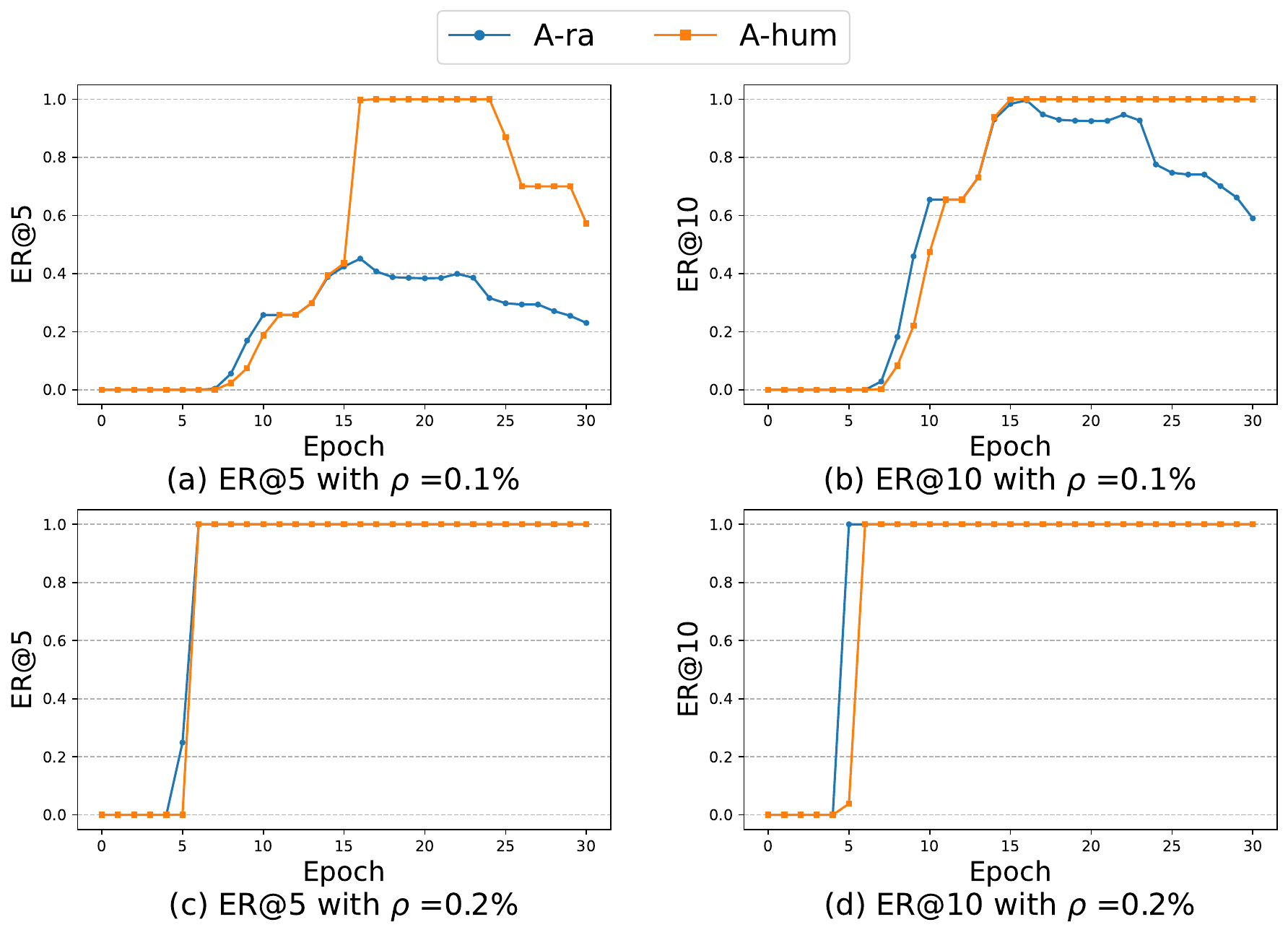}
	\caption{Attack Effectiveness under A-ra and A-hum with $0.1\%$ malicious users and $0.2\%$ malicious users.}
	\label{f3}
\end{figure}
By analyzing the effectiveness of our attacks on ML with $0.1\%$ malicious users and $0.2\%$ malicious users in Figure~\ref{f3}, it can be easily deduced that, although the strategy of hard user mining leads to a slower rise in attack effectiveness, it does improve overall attack effectiveness throughout the whole training epochs.
More specifically, we compare ER@5 and ER@10 under A-ra and A-hum on ML as following:
\textbf{(i)} As for A-hum, the values of ER@5 and ER@10 can reach maximum of 1, and respectively result in $0.5728$ and $1$ at the point that training process ends.
\textbf{(ii)} As for A-ra, maximum values of $0.4514$ and $0.9960$ in same metrics are reached, decreased to $0.2302$ and $0.5900$ at the end.
We argue that A-hum benifits from the strategy of hard user mining indeed.
The trouble brought by hard users is more influential and challenging in cases of the proportion of malicious users being extremely small, and obviously A-hum is more effective in confronting the challenge.


\section{Conclusion and Future Work}
In this paper, we aim to attack deep learning based RS in FL scenarios without attacker's prior knowledge.
We proposed two attack methods A-ra and A-hum, that use different ways to approximate benign users' embedding vectors.
Experimental results demonstrate that our attacks set the state-of-the-art, and prove that there is necessary improvement should be made in FR.
As part of future work, we will further explore the methods to detect the attacks in FR.

\section*{Acknowledgments}
This research is supported by the National Key R\&D Program of China (2021YFB2700500, 2021YFB2700502).
Specially, the authors thank Jiasheng Chen (jasonalpaca@foxmail.com) as an independent researcher for his important contribution to the codes of this work.

\newpage
\bibliographystyle{named}
\bibliography{ijcai22}
\end{document}